\documentclass[12pt,fleqn]{iopart}
\usepackage{graphicx,subfigure}
\usepackage{epstopdf}
\usepackage{color} 
\usepackage{bm}
\usepackage[]{natbib}
\bibpunct{[}{]}{;}{a}{,}{,~}
\usepackage{float} 
\usepackage{enumerate}
\usepackage{pifont}
\usepackage{cuted}
\usepackage[normalem]{ulem}
\usepackage{soul,xcolor}
\setstcolor{red}
\usepackage[linktocpage=true,hidelinks]{hyperref}

\renewcommand{\vec}[1]{{\bf #1}}
\begin{document}
\title[Benchmarking between fluid and global models ...]
{Benchmarking between fluid and global models for low-pressure oxygen DC glow discharges} 

\author{
Pedro Viegas$^{1,2}$, 
Tiago Cunha Dias$^{1}$, 
Chloé Fromentin$^{1}$,
Alexander Chukalovsky$^{3}$,
Yuri Mankelevich$^{3}$,
Olga Proshina$^{3}$,
Tatyana Rakhimova$^{3}$,
Vasco Guerra$^{1}$
and
Dmitry Voloshin$^{3}$}
\address{$^1$ Instituto de Plasmas e Fusão Nuclear, Instituto Superior Técnico - Universidade de Lisboa, Av. Rovisco Pais 1, 1049-001 Lisboa, Portugal}
\address{$^2$ Department of Physical Electronics at the Faculty of Science, Masaryk University, Kotlářská 267/2, 611 37 Brno, Czech Republic}
\address{$^3$ Skobeltsyn Institute of Nuclear Physics, Lopmonosov Moscow State University, Russia}

\ead{pedro.a.viegas@tecnico.ulisboa.pt}
\vspace{10pt}
\begin{indented}
\item[]\today
\end{indented}
\begin{abstract}
This work focuses on the benchmarking between a zero-dimensional (0D) global model (LoKI) and a one-dimensional (1D) radial fluid model for the positive column of oxygen DC glow discharges 
in a tube of 1 cm inner radius 
at pressures between 0.5 Torr and 10 Torr.
The data used in the two models are the same, so that the difference between the models is reduced to dimensionality.
A good agreement is found between the two models on the 
main discharge parameters, with relative differences below 5\%.
The agreement on species average number densities, charged and neutral, is slightly worse, with relative differences increasing with pressure from 11\% at 0.5 Torr to 57\% at 10 Torr.
The success of the 0D global model in describing these plasmas through volume averaged quantities decreases with pressure, due to pressure-driven narrowing of radial profiles. 
Hence, 
in the studied conditions, 
we recommend the use of 
volume-averaged 
models only in the pressure range up to 10 Torr.
%
\end{abstract}
%
%
\noindent{\it Keywords}:
Oxygen kinetics,
Discharge spatial resolution,
Global model,
Fluid model.
%
%
%
%
%
%
\newpage
\section{Introduction}
\label{sec:intro}

Low-temperature plasmas and gas discharges are non-equilibrium systems, where neutral and charged particles undergo collisional and radiative processes, energy transfers, transport and the effect of electromagnetic fields.
Multiphysics and multiscale phenomena take place in these media, where often distinct material phases and different elements coexist.
This makes it difficult to accurately characterize discharges and their output for applications either by measuring all the physical quantities of interest or by analytical models.
As such, numerical models based on solid physical grounds and making use of increasingly advanced computational capabilities emerge as complementary tools to experimental diagnostics.
They are necessary for discharge characterization and understanding and for predicting the behaviour of important physical and chemical quantities in different plasma systems. 
Given the wide diversity of plasma sources and conditions (pressures, excitation sources, spatial and temporal scales, working gases) in the low-temperature plasma field, different formulations and algorithms are used for numerical models.
A broad overview and description of such models is provided in the topical review `Foundations of modelling of nonequilibrium
low-temperature plasmas' by \cite{Alves18}.

One widely developed and employed class of models that provides a compromise between accuracy, versatility and computational effort is that of fluid models \citep{Feoktistov95,Kulikovsky97,Braginskiy05,Alves07,Alves09,
Laguna20,Volynets20,Vialetto22}.
These are based on the solution of the spatially and temporally dependent hydrodynamic equations for the multiple species that compose the plasma.
As such, the plasma is seen from a macroscopic point of view as a multi-fluid system of electrons, ions and neutrals in a gaseous phase.
The hydrodynamic equations are derived as moments of the Boltzmann equation, i.e. macroscopic conservation laws for the following typical quantities associated to each species: particle number, mass, momentum and energy.
These conservation equations are usually coupled to Maxwell's equations or Poisson's equation, that describes not only the electrostatic field externally applied but also the one generated by charge separation.
Fluid models can be solved in the whole three-dimensional space, or they can take physical assumptions on some dimensions and then be solved in only one or two remaining dimensions.
For instance, for some cylindrical discharges, azimuthal symmetry may be assumed, and the plasma may be considered quasi-homogeneous in the axial direction, while having steeper gradients in the radial direction.
That can leave the spatially-resolved fluid model resolution to the radial direction alone.

Another class of plasma models is that of global models, also called zero-dimensional chemical kinetics models \citep{Lee95,Ashida95,Lieberman96,Dorai03,ZDPlasKin08,Monahan08,Plasimo09,
Hurlbatt17,Guerra19,Viegas20}. 
These are spatially-averaged models, that attempt to describe the plasma as a whole.
As such, they do not describe spatial gradients within the plasma system, although they can use them as assumptions.
Global models solve the same conservation equations as fluid models, for volume-averaged quantities.
In particular, global models usually solve particle and energy rate-balance equations.
As these models are free from the computational costs of increased dimensionality, they can often include a comprehensive reaction set for dozens or hundreds of different species.
The balance equations in global models are based on source and loss terms defined by chemical reactions and by approximations to include effects of transport.
The source terms associated with electron impact reactions are usually dependent on the electron energy distribution function (EEDF), and thus global models are often coupled to electron Boltzmann equation (EBE) solvers.
With these features, global models are very useful to reproduce trends and identify the most important source and loss mechanisms of the species of interest for applications in different conditions.
It should be pointed out that zero-dimensional chemical kinetics models can also be used as local models, rather than volume-averaged global models.
This is a valid approach when local phenomena that determine species densities evolution, such as chemical reactions, are expected to have a much shorter characteristic time-scale than non-local transport phenomena, which is typically the case at near-atmospheric pressures \citep{Liu10,Lietz16,He21,Passaras21}.

When addressing plasma simulation results, namely from fluid and global models, it is important to take into account the degree of dimensionality implicit in the model and distinguish spatially-resolved from spatially-averaged quantities.
This is particularly important when comparing simulation results with experimental measurements, that can also be obtained as averages or spatially-resolved quantities, depending on the diagnostics \citep{Hofmans20,Viegas21}.
Moreover, modellers need to weight whether the insight provided by increased dimensionality compensates the higher computational cost, or potentially the lower insight into plasma kinetics, associated to it.
While fluid models can provide spatial resolution to some important quantities in the plasma system, the spatially-averaged approach of global models may be precise enough to identify the main energy transfer pathways and when studying so-called homogeneous discharges.
These are plasmas where gradients are smooth or where spatial profiles are well known, allowing to define spatial averages as representative of the plasma as a whole.
An example of discharges where this should be the case are low pressure cylindrical discharges controlled by free diffusion or by ambipolar diffusion, where the radial distribution of electron density ($n_\mathrm{e}$) follows a paraboloidal or a zero-order Bessel function of the first kind, with zero at the cylinder wall \citep{Schottky24,Parker63,Ikegami68,Durandet89,
Fridman04book,Lieberman05,Moisan12}.

In this work, we study the positive column of low-pressure cylindrical DC glow discharges, to assess under which conditions are global models precise enough, in comparison with spatially-resolved fluid models.
These discharges are chosen as test-case for being widely studied and modelled homogeneous plasma configurations \citep{Schottky24,Raizer91,Cenian94,Gordiets95,
Fridman04book,Lieberman05,Alves07,Dyatko08,
Golubovskii11,Gudmundsson17,Silva18,Volynets18,Booth19,Morillo19,AFSilva21,Naidis21}.
We further choose oxygen glow discharges as a case of interest, due to its suitability to study the kinetics of oxygen species, which undergo all the important processes taking place in other molecular gases, such as dissociation, attachment, electronic excitation and vibrational excitation.
Moreover, oxygen species are recurrently attributed importance for applications.
One example is the interest in the kinetics of atomic oxygen, which, among other applications, is crucial for product separation in plasmas for electrochemical conversion \citep{Rohnke04,Meiss08,Patel19,Chen20,Zheng22,Pandiyan22}.
In fact, kinetics in low-pressure cylindrical oxygen DC glow discharges have been recently extensively studied through both diagnostics and simulations in the pressure range 0.2-10 Torr and the current intensity range 10-40 mA \citep{Booth19,Booth20,Booth22}.
The model employed in \cite{Booth19,Booth22} is a 1D fluid model radially-resolved, developed at Moscow State University, that solves the equations for particle conservation for the different species in the plasma, together with Poisson's equation and one equation for energy conservation.

In the current work, the discharge configuration studied in \cite{Booth19,Booth20,Booth22} is examined, for a current intensity of 30 mA and pressures varying between 0.5 and 10 Torr, as a test-case to benchmark different plasma models.
The LoKI global model \citep{Tejero19,Guerra19,AFSilva21} is employed together with the 1D-radial fluid model in \cite{Booth19,Booth22}, considering the same plasma kinetics.
In section \ref{sec:methods} the two models are presented, together with a description of the discharge conditions under study.
In section \ref{sec:res}, the simulation results from both models are compared, providing a benchmark between the two models:
firstly, radial profiles of important physical quantities obtained by the fluid model are compared with the radial profiles assumed in the global model;
then, the benchmark proceeds by comparing the average values of several species densities, gas temperature and reduced electric field obtained by both models.
Finally, an overview of the differences between the modelling results is given.

\newpage
\section{Methods and conditions}
\label{sec:methods}

\subsection{Discharge set-up under study}
\label{sec:setup}

The discharge conditions examined in \cite{Booth19,Booth20,Booth22} are addressed in this work.
The set-up consists of a DC glow discharge in O$_2$, ignited in a Pyrex tube of 1.0 cm inner radius and 56 cm length.
The current intensity is kept at 30 mA and the pressure values are varied within the interval between 0.5 Torr and 10 Torr.
The cylindrical tube outer surface is kept at a constant temperature of 50ºC. 
This is guaranteed by a water/ethanol mixture flowing through an outer envelope and connected to a thermostatic bath. 
The temperature drop across the Pyrex tube wall is considered to be negligible, of less than 2 K \citep{Booth19}. 
The distance between the hollow cathode electrodes (in side-arms) is around 50.0 cm.
The anode is connected to a positive polarity high voltage power supply and the cathode is connected to the ground.
The gas flow rate is kept low, between 3 sccm for pressures below 1 Torr and 10 sccm for 2 Torr and pressures above, so as to ensure that the gas residence time ($>1$ s) is longer than the lifetime of all the active species in the discharge.
The leak rate of air into the system is small, below 0.015 sccm, corresponding to less than 0.4\% N$_2$ in the mixture in the worst case.
Hence an axially uniform plasma column is created with constant gas composition (almost pure O$_2$) in the cylindrical vessel.


\newpage
\subsection{Modelling data}
\label{sec:data}

The same reactions, cross sections and collisional probabilities are considered in both models (1D and 0D).
These are based on the works by \cite{Ivanov99,Vasiljeva04,Kovalev05,Braginskiy05,
Booth19,Booth22}.
7 charged species and 8 neutral species are described by the kinetic scheme: e, O$^-$, O$_2^-$, O$_3^-$, O$^+$, O$_2^+$, O$_4^+$, O($^3$P), O($^1$D), O($^1$S), O$_2$(X), O$_2$($a^1\Delta_g$), O$_2$($b^1\Sigma_g^+$), O$_2$(Hz) (an effective sum of the O$_2$(A$^{\prime 3}\Delta_u$,A$^3\Sigma_u^+$,c$^1\Sigma_u^-$) Herzberg states) and O$_3$.
The table of reactions is presented in \cite{Braginskiy05}.
To these, the two following reactions are added, with corresponding rate coefficients:
\begin{eqnarray}
\mathrm{O_4^+ + O_2 \rightarrow O_2^+ + O_2 + O_2}, \\
k_{\mathrm{O_4^+ + O_2}} = 3.3 \times 10^{-12} \times \left({300 \over T_\mathrm{g}}\right)^4 \times \exp{\left({-5030 \over T_\mathrm{g}}\right)} \mathrm{[m^{-3} \cdot s^{-1}]}, \\
\mathrm{O_2(b) + O(^3P) \rightarrow O_2(X) + O(^3P)}, \\
k_{\mathrm{O_2(b)+O(^3P)}} = 10^{-16} \times \exp{\left({-3700 \over T_\mathrm{g}}\right)} \mathrm{[m^{-3} \cdot s^{-1}]}. 
\end{eqnarray}
The rate coefficients $k_{\mathrm{O_4^+ + O_2}}$ and $k_{\mathrm{O_2(b)+O(^3P)}}$ are obtained from \cite{Kozlov88} and \cite{Booth22}, respectively.
Vibrations are considered in the solution of the electron Boltzmann equation (EBE), by taking into account electron impact collisions of excitation of vibrational states of the ground electronic state O$_2$(X).
However, no detailed vibrational kinetics of O$_2$(X,v) is considered in this work, 
since it is expected to have a negligible influence on the main discharge parameters assessed here: electron density, reduced electric field and dissociation degree.
It should be noticed that the benchmark can proceed as long as the two models are coherent, as is the case.


Electron kinetics is described in both models by stationary homogeneous two-term EBE solvers. 
The cross sections for electron impact are mostly taken from \cite{Lawton78} for O$_2$ collision partner and from \cite{Laher90} for O, as in \cite{Booth22}.
However, the momentum-transfer cross section for electron scattering with O has been used, as in \cite{ISToxygen}. 
Having the same electron impact cross sections and the same type of 
EBE solver, it has been verified that the same electron energy distribution functions (EEDFs) and electron impact rate coefficients are obtained in the two models.
The verification has been obtained for a wide range of $E/n_\mathrm{g}$ for pure O$_2$ and for 20\% O - 80\% O$_2$ mixtures.
Moreover, the widely used EBE solver BOLSIG+ \citep{Bolsig} has also been employed to confirm that the same results are obtained.


The loss of particles due to quenching or recombination at the wall surface is considered in both models.
These processes affect O$_2$(a), O$_2$(b), O$_2$(Hz), O($^3$P) and O($^1$D).
The effective wall loss frequency for a species j is given by \citep{Booth22}:
\begin{eqnarray}
\label{eq:nu}
\nu_\mathrm{j,wall} = \gamma_\mathrm{j} {n_\mathrm{j,nw} \over n_\mathrm{j,av}} {v_\mathrm{j,th} \over 4} {2 \over R}, \\
v_\mathrm{j,th} = \sqrt{8 k_\mathrm{B} T_\mathrm{nw} \over \pi m_\mathrm{j}},
\end{eqnarray}
where $\gamma_\mathrm{j}$, $n_\mathrm{j,nw}$, $n_\mathrm{j,av}$, $v_\mathrm{j,th}$ and $m_\mathrm{j}$ are the wall loss probability, the near-wall density, the radially-averaged density, the thermal velocity and the atomic mass of species j.
$k_\mathrm{B}$ and $T_\mathrm{nw}$ are the Boltzmann constant and the near-wall temperature and $2/R$ is the surface-to-volume ratio in cylindrical geometry.
It should be noticed that, in the absence of significant gradients in species number densities, $n_\mathrm{j,nw} \simeq n_\mathrm{j,av}$ and thus eq. \ref{eq:nu} is reduced to $\nu_\mathrm{j,wall} \simeq \gamma_\mathrm{j} {v_\mathrm{j,th} \over 2 R}$.
The values of wall loss probabilities $\gamma_\mathrm{j}$ used in this work are outlined in table \ref{tab:gamma}.

\begin{table}[!t]
\centering
\begin{tabular}{|c|c|}
\hline
 Wall reaction & $\gamma_\mathrm{j}$ \\ 
\hline
O$_2$(a) + wall $\rightarrow$ O$_2$(X) & f($p$), $3.5 \times 10^{-4}$ - $6.4 \times 10^{-4}$  \\
O$_2$(b) + wall $\rightarrow$ O$_2$(X) & 0.135  \\
O$_2$(Hz) + wall $\rightarrow$ O$_2$(X) & 1  \\
O($^3$P) + wall $\rightarrow$ 0.5 O$_2$(X) & f($p$), $6.4 \times 10^{-4}$ - $11.6 \times 10^{-4}$  \\
O($^1$D) + wall $\rightarrow$ O($^3$P) & 1  \\
\hline
\end{tabular}
\caption{Wall loss probabilities considered in the 0D and 1D models.
In the case of O$_2$(a) quenching and O($^3$P) recombination, $\gamma_\mathrm{j}$ is pressure-dependent.}
\label{tab:gamma}
\end{table}

The values of wall loss probability of O($^3$P) have been obtained from pressure-dependent loss frequency measurements in \cite{Booth19} and eq. \ref{eq:nu}.
The wall loss probabilities of O$_2$(a) and O$_2$(b) have also been calculated from experimental measurements, reported in \cite{Booth22} in the case of O$_2$(b) and not yet published in the case of O$_2$(a).
Concerning excited states O$_2$(Hz) and O($^1$D), these are assumed to fully quench on the wall surface.


Both models consider that heat exchanges in the plasma volume take place via Joule heating and heat loss by the radiation from Herzberg states (O$_2$(Hz) $\rightarrow$ O$_2$(X) + $h \nu$), 
together with heat conduction to the wall. 
Furthermore, they take as thermal data the gas molar heat capacity at constant pressure $c_p$ and the gas thermal conductivity $\lambda_\mathrm{g}$.
The thermal conductivity is assumed to be $\lambda_\mathrm{g} = 33 \times 10^{-5} \times T_\mathrm{g}^{0.78}$ J/(s$\cdot$m$\cdot$K), based on the O$_2$ experimental data from \cite{Westenberg63}, as in \cite{Booth22}.
The heat capacities of each component (O, O$_2$, O$_3$ and O$_4$), $c_{pi}(T_\mathrm{g})$, are expressed as polynomial functions: $c_{pi}(T_\mathrm{g}) = \sum_j a_{ij} T_\mathrm{g}^j, j = 0 - 4$, with $a_{ij}$ taken for temperatures between 200 K and 1000 K from the combustion thermochemical database in \cite{Burcat05}.


\newpage
\subsection{1D radial fluid model}
\label{sec:1D}


A one-dimensional 1D($r$) discharge and chemical kinetic model with the local electric field approximation for the electron energy distribution function is used.
The standard continuity equations are used to calculate the spatial distributions of all charged and neutral gas species in the radial direction:
\begin{equation}
\label{eq:1D}
{\partial n_\mathrm{i}(r,t) \over \partial r} = {1 \over r} {\partial \over \partial r} \left( r \left( D_\mathrm{i} N {\partial X_\mathrm{i} \over \partial r}\right) - q_\mathrm{i} \mu_\mathrm{i} E_r n_\mathrm{i} \right) + S_\mathrm{i}(r,t),
\end{equation}
where $N$ is the total gas number density and $n_\mathrm{i}$ and $X_\mathrm{i}$ are the concentration and mole fraction of a species indexed by “i”.
$D_\mathrm{i}$ and $\mu_\mathrm{i}$ are the diffusion and mobility coefficients of the corresponding species and $S_\mathrm{i}$ is the total rate of production-loss of a species through different reactions.
$E_r$ is the radial component of electric field, calculated from the solution of Poisson's equation:
\begin{equation}
\label{eq:poisson}
{1 \over r} {\partial \over \partial r} (r E_r) = -4 \pi \sum_\mathrm{i} q_\mathrm{i} n_\mathrm{i},
\end{equation}
where $q_\mathrm{i}$ is the charge of the corresponding particle.
The pressure and temperature dependences of the diffusion coefficients for neutral and ionic components are assumed to be: $D_\mathrm{i} \propto T^{3/2}/P$ and $D^+ \propto T^2/P$, following \cite{Raizer91}.

The electron energy distribution function (EEDF) is determined as a function of the local reduced electric field by solving the stationary homogeneous electron Boltzmann equation using the two-term approximation, including the effect of inelastic and superelastic collisions with excited states of oxygen molecules and atoms.
The electron transport coefficients ($D_e$, $\mu_e$ and the drift velocity) and the rate constants for reactions involving electrons are then calculated from the EEDF.

The boundary conditions for equations \ref{eq:1D} and \ref{eq:poisson} include the symmetry condition on the tube axis ($r = 0$) and the losses of species on the tube wall (at $r = R$, the tube radius), which occur with probabilities $\gamma_\mathrm{i}$.

The model also takes into account the heating of the gas in the discharge.
The gas temperature $T_\mathrm{g}(r,t)$ (assuming constant gas pressure) is found from the simultaneous solution of the equations for the total enthalpy of the mixture $H(r,t)$:
\begin{eqnarray}
\label{eq:H1}
{\partial H(r,t) \over \partial r} = J_z E_z - P_\mathrm{rad} + {1 \over r} {\partial \over \partial r} \left( r \lambda_\mathrm{g} {\partial T_\mathrm{g} \over \partial r} \right) + \sum_\mathrm{i} h_\mathrm{i} J_{D_\mathrm{i}} \\
\label{eq:H2}
H(r,t) = \sum_\mathrm{i} h_\mathrm{i} n_\mathrm{i} = \sum_\mathrm{i} n_\mathrm{i} \left( \int_0^T C_{p_\mathrm{i}}(T_\mathrm{g}) d T_\mathrm{g} + h_{0_\mathrm{i}} \right).
\end{eqnarray}
$J_z E_z$ is the Joule heating term ($J_z$ is the current density in the axial direction and $E_z$ is the longitudinal component of electric field); 
$P_\mathrm{rad}$ is the heat loss term by radiation from the Herzberg states;
$\lambda_\mathrm{g}$ is the gas thermal conductivity;
$J_{D_\mathrm{i}}= - {D_\mathrm{i} \over N} \nabla \left( {n_\mathrm{i} \over N} \right)$ is the heat flux for each i-th component of the mixture.
In equation \ref{eq:H2}, $C_{p_\mathrm{i}}$ and $h_{0_\mathrm{i}}$ designate the heat capacity and the enthalpy of formation of the i-th component of the mixture, correspondingly.
The values of $\lambda_\mathrm{g}$ and $C_{p_\mathrm{i}}(T_\mathrm{g})$ have been described in the previous section.
The total particle concentration $N$ as a function of the gas temperature $T_\mathrm{g}$ is determined from the ideal gas equation at constant pressure.

Equations \ref{eq:1D} - \ref{eq:H2} 
are solved using a numerical method specially developed for stiff systems of differential equations.
The EEDF is recalculated as necessary, to account changes in the local gas composition.
The time-dependent equations for the particle number densities, the EEDF$(r,t)$, the gas temperature $T_\mathrm{g}(r,t)$ and the axial electric field $E_z(t)$ (assumed to be constant across the radius) are solved self-consistently until steady-state is reached and the input discharge current intensity is matched. 
%
The calculation time to obtain the steady state solution for each case/pressure can reach several hours, depending on the relative accuracy. 
However, the code has not been optimized for computational times.
The use of implicit schemes for eq. \ref{eq:1D} and exponential schemes in the discretization of the EBE for EEDF calculation can reduce computational time if necessary.

\newpage
\subsection{0D global model LoKI}
\label{sec:0D}

The software used to assess the validity of global models in describing DC oxygen glow discharges is the LisbOn KInetics (LoKI) model. 
LoKI comprises two modules, that in this work run self-consistently coupled:
a Boltzmann solver (LoKI-B) for the electron Boltzmann equation \citep{Tejero19,Tejero21} and a Chemical solver (LoKI-C) for the global kinetic modelling of gases \citep{Guerra19,AFSilva21}.
LoKI-B is an open-source tool, licensed under the GNU general public license and freely available at https://github.com/IST-Lisbon/LoKI. 
LoKI-C is not yet freely available.

Overall, LoKI provides a chemical and transport description of plasma species in zero dimensions 
for user-defined working conditions: mixture compositions, pressure, reactor dimensions, flow rate and excitation features.
LoKI-B solves a space independent form of the two-term electron Boltzmann equation (EBE) for non-magnetised non-equilibrium low-temperature plasmas, excited by DC or HF electric fields \citep{Tejero19} or time-dependent (non-oscillatory) electric fields \citep{Tejero21}, in different gases or gas mixtures.
As such, it calculates the EEDF and macroscopic electron parameters, including collisional rate coefficients.
LoKI-C solves the system of zero-dimensional rate balance equations for all the assigned charged and neutral species in the plasma, each with volume-averaged number density $n_\mathrm{j}$, considering collisional, radiative and transport processes:
\begin{equation}
\label{eq:rate}
{\partial n_\mathrm{j} \over \partial t} = S_\mathrm{j}^{chem} + S_\mathrm{j}^{conv} + S_\mathrm{j}^{diff}.
\end{equation}
$S_\mathrm{j}^{chem}$ is the sum of the chemical source and loss terms of species j, given in this work by the collisional and radiative processes presented in section \ref{sec:data}, and $S_\mathrm{j}^{conv}$ and $S_\mathrm{j}^{diff}$ are the transport loss terms, considering axial convective transport and radial and axial diffusive transport, respectively.

The axial convection term considered in eq. \ref{eq:rate} supposes the input of O$_2$ and the output of all species j in a spatially-averaged way at the same frequency for all species, such that the number of atoms is conserved.
The convective term is dependent on the flow rate $\Gamma_\mathrm{in}$ (given in particles per second by $\Gamma$(sccm)$\times 4.47797 \times 10^{17}$) and the cylindrical chamber volume $V$ 
\citep{AFSilva21}:
\begin{eqnarray}
S_\mathrm{O_2}^{conv} = {\Gamma_\mathrm{in} \over V}, \\ 
S_\mathrm{j}^{conv} = - {n_\mathrm{j} \over n_\mathrm{g0}} {\Gamma_\mathrm{in} \over V}, 
\end{eqnarray} 
where $n_\mathrm{g,0}$ is the gas density at the beginning of the simulation.
For the low $\Gamma$ 
considered here (of a few sccm), the axial convection term has a negligible influence on the results.

The diffusion of neutral species is taken as in \cite{Guerra19}, based on the formula of \cite{Chantry87}, that takes into account partial losses of species j to the wall with a deactivation/recombination probability $\gamma_\mathrm{j}$, such that:
\begin{eqnarray}
S_\mathrm{j}^{diff} = - {n_\mathrm{j} \over \tau_\mathrm{j}^{diff}}, \\
\label{eq:diff}
\tau_\mathrm{j}^{diff} = {1 \over D_\mathrm{j}} \left[\left({\pi \over L}\right)^2+\left({J_0 \over R}\right)^2\right]^{-1} + \frac{{2 R L \over L + R}(1 - {\gamma_\mathrm{j} \over 2})}{\gamma_\mathrm{j} v_\mathrm{j,th}},
\end{eqnarray}
where $\tau_\mathrm{j}^{diff}$ is the characteristic diffusion time and $v_\mathrm{j,th}$ is the thermal velocity of species j.
$R$ and $L$ are the radius and the length of the discharge
tube, respectively, $J_0 = 2.405$ is the first zero of the zero order Bessel function and $D_\mathrm{j}$ is the diffusion coefficient of species j.
$D_\mathrm{j}$ is given by Wilke's formula for multicomponent mixtures \citep{Cheng06}, with binary diffusion coefficients calculated as in \cite{Hirschfelder64} and \cite{Guerra19}, based on Lennard-Jones binary interaction potential parameters.
The approach taken for the diffusion of neutrals tendentiously leads to uniform radial profiles of species number densities when $\gamma_\mathrm{j} \ll 1$, and to Bessel radial profiles when $\gamma_\mathrm{j} \simeq 1$ \citep{Chantry87}.

The diffusion of positive ions is described by classical ambipolar diffusion, considering the high-pressure limit of transport theories \citep{Guerra19,Phelps90}.
Indeed, in the conditions of the glow discharge positive column under study, the radial diffusion length $\Lambda$ is about an order of magnitude higher than the positive ion mean-free-path $\lambda_+$, a condition for the use of classical ambipolar diffusion.
However, a deviation from classical ambipolar diffusion is induced by the presence of negative ions.
This influence is considered in this work, taking into account the effect of several negative ions on the electron density radial profiles. 
The approach taken is based on the work in \cite{Guerra99} and is further detailed in \cite{Dias22}.

LoKI-C includes also a gas/plasma thermal model, for the self-consistent calculation of the gas temperature.
The model considers a 1D parabolic profile of gas temperature in the radial direction:
\begin{equation}
\label{eq:para}
T_\mathrm{g}(r) = T_0 - (T_0 - T_\mathrm{nw}){r^2 \over R^2}, 0 \le r \le R,
\end{equation}
where $R$ is the tube inner radius of 1 cm, $T_0$ is the peak temperature at $r=0$ and $T_\mathrm{nw}$ is the near-wall temperature at $r=R$.
The average gas temperature used in the rate coefficient calculation in the 0D model is defined as $T_\mathrm{g} \equiv T_\mathrm{g,av} = {1 \over 2} (T_0 + T_\mathrm{nw})$.
Its temporal evolution is given by the gas thermal balance equation at constant pressure \citep{Pintassilgo14}:
\begin{equation}
\label{eq:heat}
c_p n_m {\partial T_\mathrm{g} \over \partial t} = Q_\mathrm{in} - {8 \lambda_\mathrm{g} (T_\mathrm{g} - T_\mathrm{nw}) \over R^2}.
\end{equation}
In the previous equation, $c_p$ is the gas molar heat capacity at constant pressure and $\lambda_\mathrm{g}$ is the gas thermal conductivity, defined in section \ref{sec:data}.
$n_m$ is the molar density and $Q_\mathrm{in}$ is the input power transferred to gas heating per unit volume.
In this work, for coherence with the 1D model described in section \ref{sec:1D}, $Q_\mathrm{in}$ is given by a Joule heating term ($Q = \vec{J_\mathrm{e}} \cdot \vec{E}$, where $\vec{J_\mathrm{e}}$ is the current density) 
and a heat loss by the radiation from Herzberg states: O$_2$(Hz) $\rightarrow$ O$_2$(X) + $h \nu$.
%
By neglecting the temperature drop inside the Pyrex tube, a fixed inner-wall temperature $T_\mathrm{w}$  is considered as equal to the outer-wall temperature of 50ºC or 323.15 K.
It is assumed that there is a convective heat loss between the gas and the wall, near its inner surface, defined by a flux:
\begin{equation}
\label{eq:gamma1}
\Gamma_\mathrm{nw} = h_\mathrm{gas-wall} (T_\mathrm{nw} - T_\mathrm{w}),
\end{equation}
where $h_\mathrm{gas-wall}$ is the convection coefficient between the gas and the wall.
Simultaneously, the conductive heat flux in the gas near the wall, taking into account eq. \ref{eq:para}, is defined as:
\begin{equation}
\label{eq:gamma2}
\Gamma_\mathrm{nw} = \lambda_\mathrm{g} |\nabla T_\mathrm{g}|_\mathrm{nw} = {4 \lambda_\mathrm{g} \over R} (T_\mathrm{g} - T_\mathrm{nw}).
\end{equation}
Together, eqs. \ref{eq:gamma1} and \ref{eq:gamma2} define: 
\begin{equation}
\label{eq:Tnw}
T_\mathrm{nw} = \frac{{4 \lambda_\mathrm{g} \over R} T_\mathrm{g} + h_\mathrm{gas-wall} T_\mathrm{w}}{{4 \lambda_\mathrm{g} \over R} + h_\mathrm{gas-wall}}.
\end{equation}
Eqs. \ref{eq:heat} and \ref{eq:Tnw} are solved together to find the temporal evolution of $T_\mathrm{g}$.
A $h_\mathrm{gas-wall}$ of 120 J$\cdot$s$^{-1}\cdot$m$^{-2}\cdot$K$^{-1}$ has been used, as providing a good agreement with the gas temperature experimentally measured in \cite{Booth19} for the whole pressure range considered.
Further details about the thermal model solution are provided in \cite{Dias22}.

The working conditions and the plasma species described by the model are those outlined in sections \ref{sec:setup} and \ref{sec:data}.
The rate balance equation (eq. \ref{eq:rate}) is not solved for electrons, 
as in each iteration the electron density $n_\mathrm{e}$ is given to the model and kept constant. 
LoKI solves the system of equations (eqs. \ref{eq:rate}, \ref{eq:heat} and \ref{eq:Tnw}) iteratively to guarantee that the steady-state solution satisfies the total input pressure (calculated via the ideal gas law) and quasi-neutrality (same number density of negative charges and positive charges), 
which determines the self-consistent reduced electric field $E/n_\mathrm{g}$ for the used value of $n_\mathrm{e}$. 
Moreover, it is guaranteed that the electron parameters employed are retrieved from the solution of the EBE obtained for the same gas mixture and reduced electric field $E/n_\mathrm{g}$ of the final steady-state quasi-neutral solution.
Finally, the system of equations is solved iteratively for different input values of $n_\mathrm{e}$ to guarantee that, at steady-state, $I = \pi R^2 |q_\mathrm{e}| n_\mathrm{e} v_\mathrm{e}$ provides the experimental current intensity of 30 mA;
where $q_\mathrm{e}$ is the elementary charge of electrons and $v_\mathrm{e}$ is the electron drift velocity, calculated from the mobility and electric field magnitude simulated by LoKI.
The relative tolerances for the pressure, quasi neutrality, mixture composition and current intensity cycles are, respectively, $10^{-4}$, $10^{-2}$, $10^{-4}$ and $10^{-3}$.
Each of these cycles typically needs about 10 iterations to achieve such relative errors and each iteration takes a few seconds of computational time.
As such, the final solution for each case/pressure is usually found within a few minutes.
A schematic figure of the workflow of LoKI is provided in \cite{AFSilva21} (figure 1).



\newpage
\section{Results}
\label{sec:res}

\subsection{Benchmark of radial profiles}
\label{sec:rad}

\begin{figure*}[h]
\begin{centering}
\includegraphics[width=3.0in]{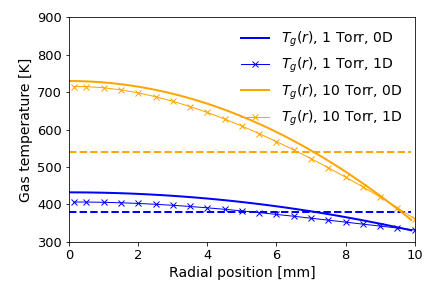}
\caption{Radial profiles of gas temperature $T_\mathrm{g}(r)$ for a lower-pressure case (1 Torr) and a higher-pressure case (10 Torr).
Profiles obtained from the 1D model and under the assumptions in the 0D model (parabolic $T_\mathrm{g}(r)$ profile between $T_\mathrm{peak}$ at $r=0$ and $T_\mathrm{nw}$ at $r=10$ mm).
The dashed lines represent the average values used in the 0D model.}
\label{Fig:Tgr}
\end{centering}
\end{figure*}

The benchmarking of 0D and 1D models starts by verifying that the radial profiles of the main physical quantities assumed in the 0D model match those obtained in 1D for the conditions under study.
One of the main parameters determining the physics in the glow discharge is the translational temperature of neutral atoms and molecules $T_\mathrm{g}$, also called gas temperature.
This is assumed in the 0D formulation to follow a parabolic profile (eq. \ref{eq:para}).
Figure \ref{Fig:Tgr} shows that this assumption is valid for a lower-pressure case (1 Torr) and for a higher-pressure case (10 Torr).
This figure represents $T_\mathrm{g}(r)$ obtained from the 1D simulation results and from the 0D assumption, that uses $T_\mathrm{g,av}$ (also represented in figure \ref{Fig:Tgr}) and $T_\mathrm{nw}$ calculated by LoKI.

It can be noticed in figure \ref{Fig:Tgr} that the 1D result of $T_\mathrm{g}(r)$, resulting from the 1D resolution of the heat equation, stands slightly below the 0D profile and is slightly flatter.
However, the difference is lower than 10\% and the 1D profile is very close to a parabola.
Another feature that can be noticed is that $T_\mathrm{g}(r)$ is more concave for higher pressure, with a larger difference between $T_\mathrm{nw}$ and the peak $T_0$.
This is determined by the increased collisionality with pressure (higher $Q_\mathrm{in}$ in eq. \ref{eq:heat}) that increases $T_\mathrm{g,av}$ and $T_0$, while $T_\mathrm{nw}$ is kept low due to wall colling and finite thermal conductivity.
Finally, it should be taken into account that, even though the 0D model assumes a parabola, only the average value $T_\mathrm{g,av}$ is actually used in rate coefficient calculations for volume reactions.
It could be more accurate for 0D models to consider the whole parabolic evolution of $T_\mathrm{g}(r)$ in the calculation of average rate coefficients, i.e. $k_\mathrm{av}(T_\mathrm{g}(r))$, instead of $k_\mathrm{av}(T_\mathrm{g,av})$.
However, we show in this work that the use of $T_\mathrm{g,av}$ provides reasonable agreement with 1D simulations.

\begin{figure*}[h]
\begin{centering}
\includegraphics[width=3.0in]{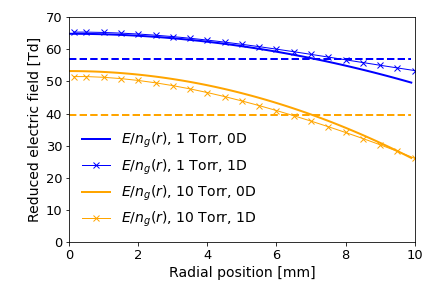}
\caption{Radial profiles of reduced electric field $E/n_\mathrm{g}(r)$ for a lower-pressure case (1 Torr) and a higher-pressure case (10 Torr).
Profiles obtained from the 1D model and under the assumptions in the 0D model (parabolic $E/n_\mathrm{g}(r)$, assuming uniform $E(r)$ and parabolic $n_\mathrm{g}(r)$ following the $T_\mathrm{g}(r)$ profile and the ideal gas law).
The dashed lines represent the average values used in the 0D model.}
\label{Fig:ENr}
\end{centering}
\end{figure*}

\newpage
The profile of $T_\mathrm{g}(r)$ is directly related to that of the gas density $n_\mathrm{g}(r)$ via the ideal gas law.
As such, it is determinant also for the main parameter accelerating electrons and determining electron impact rate coefficients, which is $E/n_\mathrm{g}(r)$.
If the axial electric field magnitude $E$ is assumed to be radially uniform and $T_\mathrm{g}(r)$ is assumed to be parabolic, then $E/n_\mathrm{g}(r)$ should also follow a parabolic profile.
This assumption is verified in figure \ref{Fig:ENr}, for the same pressures studied in the previous figure, by comparing the assumed parabolic profiles of the 0D model with the 1D simulation results.
Once more, the 1D radial profiles match closely the parabolic assumption and a more concave profile of $E/n_\mathrm{g}(r)$ is noticeable for higher pressure.
Moreover, it should be noticed again that the average value $E/n_\mathrm{g,av}$ is used to compute average rate coefficients in the 0D model, and not the parabolic profile $E/n_\mathrm{g}(r)$.

\begin{figure*}[h]
\begin{centering}
\includegraphics[width=3.0in]{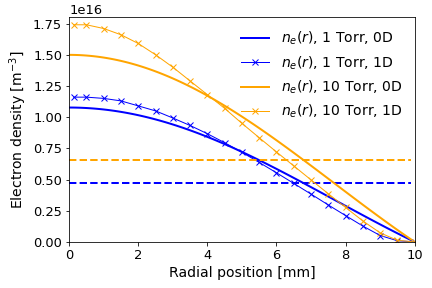}
\caption{Radial profiles of electron density $n_\mathrm{e}(r)$ for a lower-pressure case (1 Torr) and a higher-pressure case (10 Torr).
Profiles obtained from the 1D model and under the assumptions in the 0D model (Bessel $n_\mathrm{e}(r)$ profile).
The dashed lines represent the average values used in the 0D model.}
\label{Fig:ner}
\end{centering}
\end{figure*}

Another main discharge parameter is the electron density $n_\mathrm{e}$.
In homogeneous low pressure cylindrical electropositive discharges where charged particle balance is controlled by electron impact ionization and ambipolar diffusion, $n_\mathrm{e}$ follows a zero-order Bessel function of the first kind, with zero at the cylinder wall \citep{Schottky24,Parker63,Ikegami68,Durandet89,
Fridman04book,Lieberman05,Moisan12}.
The presence of negative ions induces a deviation from this profile \citep{Gousset89,Guerra99,Dias22}.
In this work, for the sake of simplicity, we test the 1D simulation results against a Bessel assumption in figure \ref{Fig:ner}. 
While the agreement is good for the 1 Torr case, it decreases with pressure, and relative differences of 17\% can be observed for the peak value of $n_\mathrm{e}(r)$ at 10 Torr.


The increasing deviation of $n_\mathrm{e}(r)$ from the Bessel profile with 
a change in plasma parameters (pressure, current) can be caused by the development of instabilities leading to discharge stratification and further contraction.
The radial contraction of plasmas has been reported in many works \citep{Kenty62,Petrov99,Kabouzi02,Dyatko08,Shkurenkov09,
Golubovskii11,Wolf19,Viegas20,Viegas21,Vialetto22}.
The narrowing of the $n_\mathrm{e}$ profile observed in figure \ref{Fig:ner} is not due to a departure from the ionization-diffusion regime, since in the current conditions and considering the chosen reaction scheme, diffusive losses are still the dominant charge loss process.
In addition, we emphasize once more that all the studied discharge regimes in the present paper correspond to experimental conditions with a visible uniform radial discharge distribution, 
without striations \citep{Booth19,Booth20,Booth22}.
Furthermore, the contraction is not attributed to electronegativity, since according to the 1D simulation results, an increase of 
attachment in the center of the plasma, on its own, leads to a flattening of $n_\mathrm{e}(r)$ instead of a contraction. 
We attribute the pressure-driven change of $n_\mathrm{e}$ radial profile in 
the positive column of the oxygen glow discharge to inhomogeneous gas heating, which can lead to the so-called thermal-ionization instability with increasing pressure and current, as in other studies of plasma contraction \citep{Martinez04,Moisan12,Shneider12,Shneider14,Ridenti18,Zhong19}.
Indeed, we have observed in figure \ref{Fig:ENr} that $E/n_\mathrm{g}(r)$ is concave and therefore the electron impact ionization rate coefficient 
$k_i$ ($k_i = {\mathrm{[O_2]} \over n_\mathrm{g}} k_{\mathrm{O_2 \rightarrow O_2^+}} + {\mathrm{[O]} \over n_\mathrm{g}} k_{\mathrm{O \rightarrow O^+}}$) 
 also has a concave profile.
As a result, the assumption of radial homogeneity leading to a Bessel $n_\mathrm{e}(r)$ profile is not generally satisfied.
For the lower-pressure case of 1 Torr, the deviation of $E/n_\mathrm{g}(r)$ from uniformity is small ($E/n_\mathrm{g}$ varies only between 53 and 65 Td in the 1D simulation results) and hence the $n_\mathrm{e}(r)$ profile is very close to a Bessel profile.
Nevertheless, the inhomogeneous gas heating observed in figure \ref{Fig:Tgr} increases with pressure and leads to a sharper gradient of $E/n_\mathrm{g}(r)$ at 10 Torr, with values from the 1D results in figure \ref{Fig:ENr} varying between 26 and 51 Td.
%
Indeed, according to the 1D simulation results, while at 1 Torr $k_i$ varies only between around $2.4 \times 10^{-18}$ m$^3\cdot$s$^{-1}$ at $r=0$ and $5.4 \times 10^{-19}$ m$^3\cdot$s$^{-1}$ at $r=R$, at 10 Torr the variation is much wider, between around $2.3 \times 10^{-19}$ m$^3\cdot$s$^{-1}$ and $8.6 \times 10^{-23}$ m$^3\cdot$s$^{-1}$.
%
It should be noticed that the increased ionization in the plasma core with respect to the plasma edges contributes to increased $T_\mathrm{g}$ in the core via Joule heating, thus forming a self-reinforcing cycle between heating and ionization, that can finally lead to the radial contraction \citep{Zhong19}.
The radial contraction cannot be captured by the 0D model, that always considers $n_\mathrm{e,av}$, based on the low radial variation of $n_\mathrm{e}(r)$ under the Bessel assumption.

\begin{figure*}[h]
\begin{centering}
\subfigure[\label{Fig:Or1}]{\includegraphics[width=3.0in]{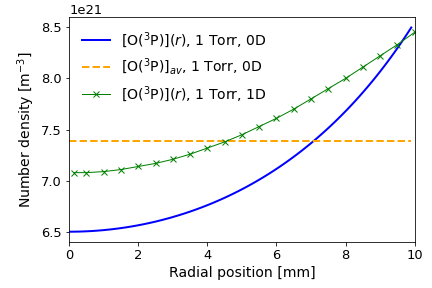}}
\subfigure[\label{Fig:Or10}]{\includegraphics[width=3.0in]{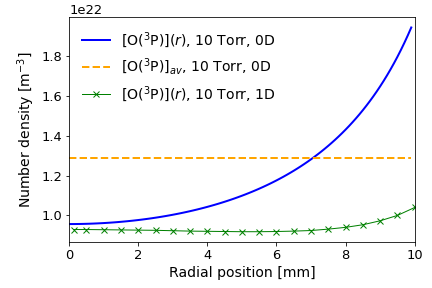}}
\caption{Radial profiles of ground-state atomic oxygen density [O($^3$P)]($r$) for a lower-pressure case (1 Torr) (a) and a higher-pressure case (10 Torr) (b).
Profiles obtained from the 1D model and under the assumptions in the 0D model (uniform [O($^3$P)]$_\mathrm{av}$ or following the parabolic $T_\mathrm{g}(r)$ profile: [O($^3$P)]($r$)=[O($^3$P)]$_\mathrm{av} \times T_\mathrm{g,av}/T_\mathrm{g}(r)$).}
\label{Fig:Or}
\end{centering}
\end{figure*}

In the study of oxygen glow discharges, the spatial distribution of the number density of atomic oxygen is also of paramount importance, due to the high reactivity of this species.
The radial profiles of O($^3$P) density are represented in figure \ref{Fig:Or}, for 1 Torr (figure \ref{Fig:Or1}) and for 10 Torr (figure \ref{Fig:Or10}).
The 1D simulation result is presented, together with the 0D-calculated average density and the profile obtained from it by assuming proportionality with the total gas density, which, according to the ideal gas law, gives:
[O($^3$P)]($r$)=[O($^3$P)]$_\mathrm{av} \times T_\mathrm{g,av}/T_\mathrm{g}(r)$).
This assumption is based on an assumed radial uniformity of source (electron-impact dissociation) and loss (diffusion and recombination at the surface and in volume) processes of atomic oxygen, 
consistent with $\gamma_\mathrm{j} \ll 1$ (see table \ref{tab:gamma}) in eq. \ref{eq:diff}. 

Figure \ref{Fig:Or} shows that, for both pressures, the 1D and 0D models provide O($^3$P) densities within the same order of magnitude but with slightly different spatial evolutions.
For 1 Torr, the 1D simulation result approximately follows the ideal gas law, that provides a difference between minimum and maximum values below a factor 2.
The 0D parabolic profile provides a good agreement near the wall but slightly lower values in the bulk of the plasma.
However, for 10 Torr, the 0D parabolic profile provides a similar result to 1D in the bulk but a factor 2 higher near the wall.
Indeed, at 10 Torr, the 1D radial profile of O($^3$P) density is almost flat, rather than parabolic.
The reasons for not following the ideal gas law are related with diffusivity, O atom wall losses and the radial non-uniformity of $E/n_\mathrm{g}(r)$ (see figure \ref{Fig:ENr}) and thus of the electron impact dissociation rate coefficient. 
It has been shown in \cite{Booth19} that oxygen atoms in a similar DC discharge conditions are predominantly lost by recombination on the tube surface and hence the assumption of radially uniform losses is not fulfilled.

As a result of the previous analysis, we consider that, when using a 0D model without access to the resolved spatial distribution of O($^3$P) density, it is generally more accurate to consider flat profiles than parabolic ones.
This consideration has implications on the calculation of source and loss terms in the rate balance equations (eq. \ref{eq:rate}) and on the calculation of wall loss probabilities from wall loss frequency measurements (eq. \ref{eq:nu}) by considering [O($^3$P)]$_\mathrm{nw}$=[O($^3$P)]$_\mathrm{av}$. 


In this section, it has been shown that the radial profiles of some of the main discharge parameters obtained through 1D simulation results in the studied conditions are not far from those assumed in 0D models:
$T_\mathrm{g}(r)$ and $E/n_\mathrm{g}(r)$ are approximately parabolic, $n_\mathrm{e}(r)$ are near-Bessel and [O($^3$P)]$(r)$ are almost flat.
The deviation from these assumptions generally increases with pressure, due to the pressure-driven narrowing of radial profiles, 
but is still rather small (below a factor 2) up to 10 Torr in the positive column of oxygen glow discharges.
This shows that, even though the 1D model takes the full radially-resolved profiles and the 0D model uses only averaged quantities, it is valid to compare the averaged values from both models, as they refer to the same type of radial profiles. 
That type of comparison is the subject of the next section.


\newpage
\subsection{Benchmark of average values}
\label{sec:av}

\begin{figure*}[h]
\begin{centering}
\subfigure[\label{Fig:Tsp}]{\includegraphics[width=3.0in]{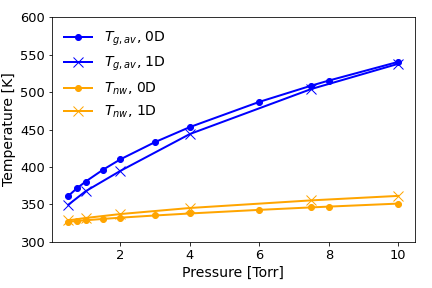}}
\subfigure[\label{Fig:ENp}]{\includegraphics[width=3.0in]{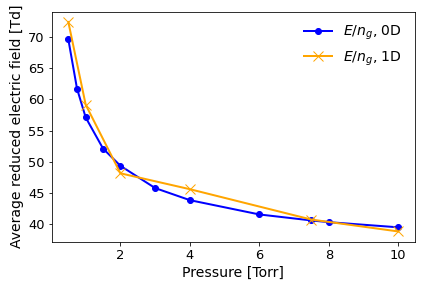}}
\caption{Average gas temperature $T_\mathrm{g,av}$ and near-wall temperature $T_\mathrm{nw}$ (a) and average reduced electric field $E/n_\mathrm{g}$ (b), as $f(p)$, from 1D and 0D models.}
\label{Fig:Tp}
\end{centering}
\end{figure*}

To proceed with the benchmarking, $T_\mathrm{g,av}$, $T_\mathrm{nw}$ and the average values of $E/n_\mathrm{g}$ are compared between the 0D and 1D simulation results for the whole range of pressures between 0.5 Torr and 10 Torr, in figure \ref{Fig:Tp}. 
It can be noticed that all the values are very close, with differences below 10\%.
$T_\mathrm{g,av}$ obtained from the 1D model is always slightly lower than the one from the 0D calculations, while $T_\mathrm{nw}$ is slightly higher in the 1D simulations, which reflects the flatter radial profiles of $T_\mathrm{g}(r)$ in 1D results observed in figure \ref{Fig:Tgr}.
$E/n_\mathrm{g}(p)$ follows the same evolution with pressure in both models, with maxima below 75 Td at 0.5 Torr and minima above 35 Td at 10 Torr.
The electron temperature $T_\mathrm{e}(p)$ (not represented here) presents a similar profile, with maxima at around 2.7 eV and minima near 2.1 eV.

\begin{figure*}[h]
\begin{centering}
\includegraphics[width=3.0in]{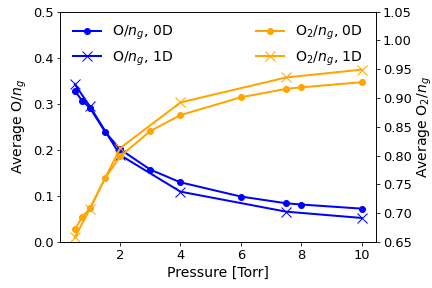}
\caption{Average molar fractions of diatomic O$_2$ ([O$_2]/n_\mathrm{g}$) and atomic O ([O]$/n_\mathrm{g}$) as $f(p)$, from 1D and 0D models.
}
\label{Fig:O2p}
\end{centering}
\end{figure*}

The comparison proceeds with the dissociation degree.
Figure \ref{Fig:O2p} shows the average molar fractions of O and O$_2$ for the several pressures.
The calculation of the molar fractions includes all the O and O$_2$ neutral and ionized species considered in the models.
The agreement between 0D and 1D models is good but decreases with pressure.
Indeed, at higher pressures, the plasma is slightly (a few percent) more dissociated in the 0D model than in the 1D case, as already suggested by the results in figure \ref{Fig:Or}.
This small difference is justified mostly by the particle balance of O($^3$P), where electron impact dissociation of O$_2$(X) and O$_2$(a) are the main source terms, and volume and wall recombination are the main loss terms.
As the $E/n_\mathrm{g}$ profile is concave and the O($^3$P), O$_2$(X) and O$_2$(a) profiles are convex in the 1D simulation results at 10 Torr (see figures \ref{Fig:ENr}, \ref{Fig:Or} and \ref{Fig:O3}), dissociation is hindered and recombination is promoted, relatively to the 0D results that only consider average values in rate calculations.

\begin{figure*}[h]
\begin{centering}
\subfigure[\label{Fig:chargep}]{\includegraphics[width=3.0in]{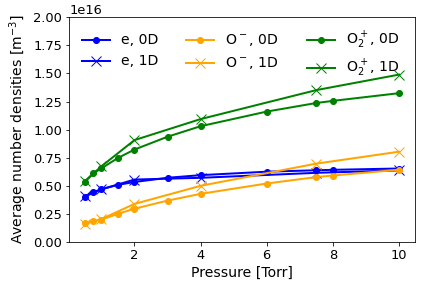}}
\subfigure[\label{Fig:charger}]{\includegraphics[width=3.0in]{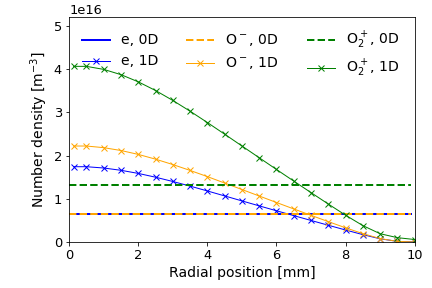}}
\caption{(a) Average number densities of main charged particles as $f(p)$, from 1D and 0D models: e, O$^-$ and O$_2^+$. 
(b) Radial profiles of the same densities, at $p = 10$ Torr.
}
\label{Fig:charges}
\end{centering}
\end{figure*}

\newpage
The agreement between 0D and 1D simulation results continues when we analyse the average number densities of charged species.
Figure \ref{Fig:charges} shows only the main charged species: 
electrons, O$^-$ and O$_2^+$; 
since the densities of the remaining ions (O$_2^-$, O$_3^-$, O$^+$ and O$_4^+$) are negligible.
While the average electron density always presents a good agreement between 0D and 1D simulation results, the plasma contains more (up to 25\%) negative and positive ion densities in the 1D model than in 0D.
The difference increases with pressure and is attributed to the concave profiles of $n_\mathrm{e}$ and $E/n_\mathrm{g}$ (see figures \ref{Fig:ENr} and \ref{Fig:ner}) that promote the production of negative and positive ions via electron impact reactions (attachment and ionization) in 1D simulations with respect to the 0D model, that considers only averages.
Concerning the destruction of O$^-$ and O$_2^+$, it takes place mostly via detachment by collisions with species with convex radial profiles (O$^-$ case) and via wall recombination (O$_2^+$ case).
As these species have concave radial profiles (see figure \ref{Fig:charger}), their destruction is relatively decreased in 1D, which also contributes to the observed higher average densities.

\begin{figure*}[h]
\begin{centering}
\includegraphics[width=3.0in]{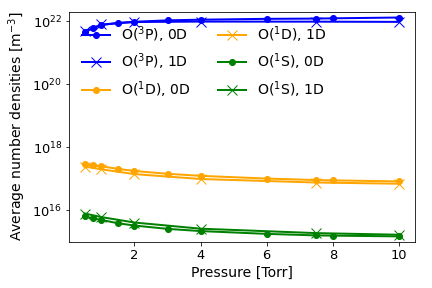}
\caption{Average atomic excited state densities as $f(p)$, from 1D and 0D models: O($^3$P), O($^1$D) and O($^1$S).
}
\label{Fig:atomp}
\end{centering}
\end{figure*}

\begin{figure*}[h]
\begin{centering}
\subfigure[\label{Fig:O2mol}]{\includegraphics[width=3.0in]{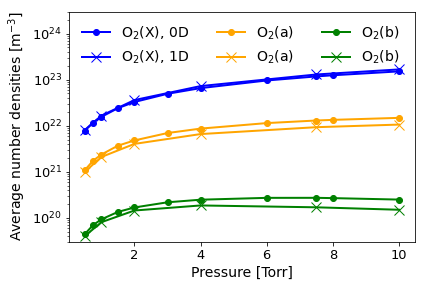}}
\subfigure[\label{Fig:O3mol}]{\includegraphics[width=3.0in]{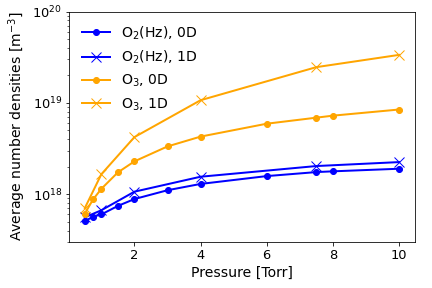}}
\caption{Average molecular excited state number densities as $f(p)$, from 1D and 0D models: O$_2$(X), O$_2$(a) and O$_2$(b) (a); O$_2$(Hz) and O$_3$ (b).
}
\label{Fig:molp}
\end{centering}
\end{figure*}

\begin{figure*}[h]
\begin{centering}
\subfigure[\label{Fig:O3prod}]{\includegraphics[width=3.0in]{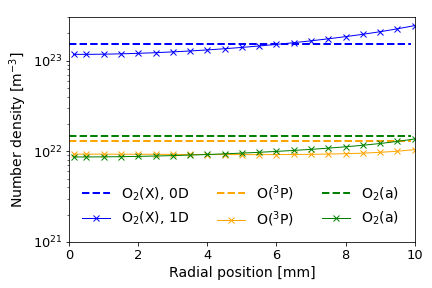}}
\subfigure[\label{Fig:O3dest}]{\includegraphics[width=3.0in]{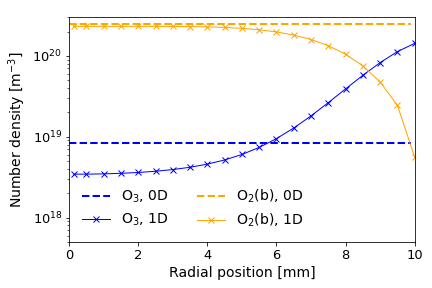}}
\caption{Radial profiles of the number densities of the reactants taking part in the main source reactions and loss reactions of O$_3$ at 10 Torr.
Profiles from the 1D model (full line with $\times$ marker) and uniform values used in the 0D model (dashed lines).
}
\label{Fig:O3}
\end{centering}
\end{figure*}

Concerning excited state average densities, these are represented from 0D and 1D simulation results in figures \ref{Fig:atomp} and \ref{Fig:molp}.
Despite the logarithmic vertical scale in the figures, it can be noticed that the agreement between the two models is quite good for most species.
However, there is significantly more (a factor 3 at 10 Torr) O$_3$ average density in the 1D simulation results than in the 0D case.
The deviation increases with pressure.
It is important to assess this difference.

\newpage
The main reactions of production of O$_3$ at the higher pressures, e.g. 10 Torr, and their rate coefficients, are:
\begin{eqnarray}
\mathrm{O(^3P) + O_2(X) + O_2(X) \rightarrow O_3 + O_2(X)}, \\
k_\mathrm{O(^3P) + 2 O_2(X)} = 5.6 \times 10^{-41} \times T_\mathrm{g}^{-2} [\mathrm{m^6 \cdot s^{-1}}], \\
\mathrm{O^- + O_2(a) \rightarrow e + O_3}, \\
k_\mathrm{O^- + O_2(a)} = 1.9 \times 10^{-16} [\mathrm{m^3 \cdot s^{-1}}], \\
\mathrm{O_2(X) + O(^3P) + O(^3P) \rightarrow O_3 + O(^3P)}, \\
k_\mathrm{O_2(X) + 2 O(^3P)} = 2.15 \times 10^{-46} \times \exp{(345/T_\mathrm{g})} [\mathrm{m^6 \cdot s^{-1}}].
\end{eqnarray}

On average, the 1D results at 10 Torr contain slightly more O$^-$ and O$_2$(X) than the 0D results, which promotes ozone production.
Conversely, the 1D results contain slightly less O$_2$(a) and O($^3$P) averaged densities, which hinders ozone production.
Therefore, an analysis of the average densities of reactants is inconclusive in terms of justifying the O$_3$ density difference.
Nevertheless, we should notice that the rate coefficients for O$_3$ production decrease with $T_\mathrm{g}$.
Despite similar $T_\mathrm{g,av}$, $T_\mathrm{g}$ is lower on the edges in the 1D model than in 0D, where only $T_\mathrm{g,av}$ is used (see figure \ref{Fig:Tgr}).
This is also the region where the reactants have higher number densities, as is shown in figure \ref{Fig:O3}, where 1D radial profiles are compared with the averages used in the 0D model at 10 Torr. 
Together, these two factors determine higher source of O$_3$ in the 1D model than in 0D.


Moreover, regarding the losses of O$_3$, we should notice that at 10 Torr the main reactions of destruction of O$_3$ and their rate coefficients are:
\begin{eqnarray}
\mathrm{O_2(a) + O_3 \rightarrow O(^3P) + O_2(X) + O_2(X)}, \\
k_\mathrm{O_2(a) + O_3} = 5.2 \times 10^{-17} \times \exp{(-2840/T_\mathrm{g})} [\mathrm{m^3 \cdot s^{-1}}], \\
\mathrm{O_2(b) + O_3 \rightarrow O(^3P) + O_2(X) + O_2(X)}, \\
k_\mathrm{O_2(b) + O_3} = 1.5 \times 10^{-17} [\mathrm{m^3 \cdot s^{-1}}], \\
\mathrm{O(^3P) + O_3 \rightarrow O_2(X) + O_2(X)}, \\
k_\mathrm{O(^3P) + O_3} = 7.68 \times 10^{-18} \times \exp{(-2060/T_\mathrm{g})} [\mathrm{m^3 \cdot s^{-1}}].
\end{eqnarray}

The reactants for these loss processes (except O$_3$ itself), O$_2$(a), O$_2$(b) and O($^3$P), have slightly lower average densities at high pressures in the 1D simulation results than in the 0D case (see figures \ref{Fig:atomp} and \ref{Fig:molp}).
Furthermore, the rate coefficients for O$_3$ destruction increase with $T_\mathrm{g}$, and thus are lower in the plasma edges, where most reactants, O$_2$(a), O($^3$P) and especially O$_3$, have higher densities (see figure \ref{Fig:O3}).
These two factors contribute to a relatively lower destruction of O$_3$ in the 1D model.
We can conclude that the subtle differences in $T_\mathrm{g}$ and species densities induced by dimensionality  lead to both higher production and relatively lower destruction of O$_3$ in the 1D model than in the 0D case.
The difference could potentially be decreased by using parabolic profiles of $T_\mathrm{g}(r)$ and $E/n_\mathrm{g}(r)$ 
in the 0D model rate coefficient calculations, instead of average values, i.e. by using $k_\mathrm{av}(T_\mathrm{g}(r),E/n_\mathrm{g}(r))$, instead of $k_\mathrm{av}(T_\mathrm{g,av},E/n_\mathrm{g,av})$.


\begin{figure*}[h]
\begin{centering}
\includegraphics[width=3.0in]{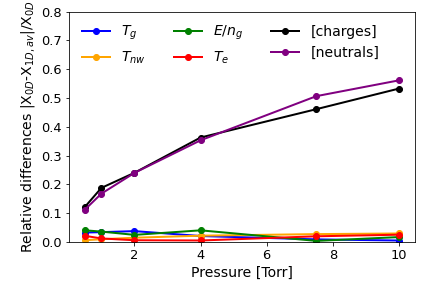}
\caption{Relative differences between 0D and 1D averaged results, as function of pressure, for $R=1.0$ cm.
Concerning species densities, the averages of the relative differences of all charged species densities and all neutral species densities are considered.}
\label{Fig:diff}
\end{centering}
\end{figure*}

\newpage
The differences found between 0D and 1D averaged simulation results are summarized in figure \ref{Fig:diff}.
This figure represents, as a function of pressure, the relative differences between the values found for temperatures ($T_\mathrm{g}$, $T_\mathrm{nw}$ and $T_\mathrm{e}$), reduced electric field ($E/n_\mathrm{g}$) and species densities.
Concerning species densities, the averages of the relative differences of all charged species densities and all neutral species densities are considered
(i.e. the relative differences are summed and the sum is divided by the number of species in question).

\newpage
Figure \ref{Fig:diff} shows that, in the whole pressure range between 0.5 Torr and 10 Torr, there is a very good agreement (with differences below 5\%) on average temperatures and reduced electric field obtained from 0D and 1D models.
The disagreement on average species densities is also relatively low, but it increases with pressure.
Indeed, the relative differences lie below 20\% at 1 Torr and between 50\% and 60\% at 10 Torr.
The fundamental reason for this increase appears to be the pressure-driven narrowing of radial profiles. 
As pressure increases, radial profiles become more concave and radial gradients become sharper in the 1D description.
As a result, the success of 0D models in describing oxygen glow discharges through volume averaged quantities decreases with pressure.
In the case under study, with 1 cm tube inner radius, in light of the results, we would recommend the use of 0D global models to describe this plasma in the pressure range up to 10 Torr, but not above, since the deviation with respect to radially-resolved results is expected to increase.
For lower tube radii, as the discharge parameters are expected to be less radially uniform, the upper limit of pressure of validity of global models may be lower.
Likewise, for higher tube radii, it may be higher. 

\newpage
\section{Conclusions}
\label{sec:conc}

This work has focused on the benchmarking between different models for the positive column of low-pressure oxygen DC glow discharges.
The models compared are a zero-dimensional (0D) global model, LoKI \citep{Tejero19,Guerra19,AFSilva21}, and a one-dimensional (1D) radial fluid model \citep{Booth19,Booth22}.
In the cylindrical low-pressure glow discharges under study, azimuthal symmetry is assumed and the plasma is considered quasi-homogeneous in the axial direction, leaving the need for its description to the radial direction alone.
In this manuscript we have assessed under which conditions is the spatially-averaged approach of global models precise enough in comparison with the radially-resolved fluid model.
The discharge configuration examined is the one in \cite{Booth19}, for a current intensity of 30 mA and pressures varying between 0.5 and 10 Torr.
The data used in the two models (reaction scheme, electron impact cross sections, rate coefficients and thermal data) and the electron Boltzmann equation solvers employed are the same, so that the difference between the models is reduced to dimensionality.

A good agreement has been found between the two models on the values and radial profiles of the main discharge parameters.
Indeed, the gas temperature profiles $T_\mathrm{g}(r)$ obtained in the 1D simulation results are parabolic, as supposed in the global model.
The same applies to the parabolic profile of reduced electric field $E/n_\mathrm{g}(r)$ and to the Bessel profile of electron density $n_\mathrm{e}(r)$.
These profiles become more concave as pressure increases, 
and $n_\mathrm{e}(r)$ starts deviating from a Bessel profile at 10 Torr.
Although the shape of these profiles is assumed in the global model formulation, only the average values $T_\mathrm{g,av}$, $E/n_\mathrm{g,av}$ and $n_\mathrm{e,av}$ are used in the source term calculations in 0D.
This is a main difference of dimensionality that is aggravated as pressure increases.

The average values of $T_\mathrm{g}$, $n_\mathrm{e}$, $E/n_\mathrm{g}$, electron temperature $T_\mathrm{e}$, near-wall temperature $T_\mathrm{nw}$ and dissociation fraction found with the two models are very similar for the whole pressure range considered, with relative differences below 5\%.
The agreement on species average number densities, charged and neutral, is slightly worse, with relative differences increasing with pressure from 11\% at 0.5 Torr to 57\% at 10 Torr.
However, the results always have the same order of magnitude, with the only case of visible discrepancy (up to a factor 3) being the average number density of O$_3$.
That deviation has been attributed mostly to the concave profile of $T_\mathrm{g}(r)$, that is not considered in the 0D model.

The results analysed show that the success of 0D global models in describing the positive column of oxygen glow discharges through volume averaged quantities decreases with pressure, due to pressure-driven narrowing of radial profiles. 
Hence, 
for the studied conditions with 1 cm tube inner radius, we would recommend their use only in the pressure range up to 10 Torr.
Nevertheless, we point out that 0D models can be used as local models at higher pressures, in conditions where the characteristic time-scales of local reactive phenomena and species densities evolution are much shorter than those of non-local transport. 

\newpage
\ack
This work was supported by the Portuguese FCT - Fundaç\~ao para a Ciência e a Tecnologia, under projects UIDB/50010/2020, UIDP/50010/2020 and PTDC/FIS-PLA/1616/2021 (PARADiSE).
PV acknowledges support by project LM2018097 funded by the Ministry of Education, Youth and Sports of the Czech Republic.
Moreover, this work was partially supported by the European Union's Horizon 2020 research and innovation programme under grant agreement MSCA ITN 813393. \\
Part of the study related to 1D modelling was supported by the Russian Science Foundation, under project no. 21-72-10040.
T. Rakhimova, and D. Voloshin would like to acknowledge the Interdisciplinary Scientific and Educational School of Moscow University ``Photonic and Quantum Technologies - Digital Medicine".
A. Chukalovsky and Yu. Mankelevich would like to acknowledge the Interdisciplinary Scientific and Educational School of Moscow University ``Fundamental and applied space research".

\newpage

\bibliographystyle{agufull04}
\bibliography{Paperbib}
%
%
\end{document}